\documentclass{iaus}
\usepackage{graphicx}
\usepackage{subfigure}
\usepackage{amsmath}
\usepackage{natbib}
\usepackage{bm}

\title[3D MHD simulations of subsurface convection]  
{3D MHD simulations of subsurface convection in OB stars}

\author[Matteo  Cantiello et al.]   
{Matteo  Cantiello$^1$, Jonathan Braithwaite$^1$, Axel Brandenburg$^{2,3}$, Fabio Del Sordo$^{2,3}$, Petri K\"apyl\"a$^{2,4}$, Norbert Langer$^1$}
 
\affiliation{$^1$Argelander-Institut f\"ur Astronomie der Universit\"at Bonn, Auf dem H\"ugel 71, D--53121 Bonn, Germany
 email: {\tt cantiello@astro.uni-bonn.de} 
$^2$NORDITA, AlbaNova University Center, Roslagstullsbacken 23, SE-10691 Stockholm, Sweden
$^3$Department of Astronomy, AlbaNova University Center,
Stockholm University, SE--10691 Stockholm, Sweden 
$^4$Department of Physics, Gustaf H\"allstr\"omin katu 2a (PO Box 64), FI-00014, University of Helsinki, Finland}
 
\pubyear{2010}
\volume{272}   
\pagerange{119--126}
\jname{Active OB stars}
\editors{C. Neiner, G. Wade, G. Meynet \& G. Peters, eds.}

\def\aap{\mbox{\it{Astron.\ Astrophys.}}}

\def\apjl{\mbox{\it{Astrophys.\ J.}}}

\def\apjs{\mbox{\it{Astrophys.\ J.\ Suppl.}}}

 \def\simle{\mathrel{\hbox{\rlap{\hbox{\lower4pt\hbox{$\sim$}}}\hbox{$<$}}}}
 \def\simgr{\mathrel{\hbox{\rlap{\hbox{\lower4pt\hbox{$\sim$}}}\hbox{$>$}}}}

 \def\c2{^{12}{\rm C}}
 \def\c3{^{13}{\rm C}}
 \def\n14{^{14}{\rm N}}
 \def\c1213{^{12}{\rm C}/^{13}{\rm C}}
 \def\he3he4{^3{\rm He}/^4{\rm He}}

\newcommand{\EQ}{\begin{equation}}
\newcommand{\EE}{\end{equation}}
\newcommand{\EQA}{\begin{eqnarray}}
\newcommand{\EEA}{\end{eqnarray}}

\newcommand{\pd}{\partial}

\newcommand{\DIV}{\bm{\nabla} \cdot }

\def\onethird{{\textstyle{1\over3}}}
\def\onehalf{{\textstyle{1\over2}}}

\begin{document}

\maketitle

\begin{abstract}
During their main sequence evolution, massive stars can develop convective regions very close to  their surface.
These regions are caused by an opacity peak associated with iron
ionization. Cantiello et al.\ (2009) found a possible connection between
the presence of sub-photospheric convective motions and small scale
stochastic velocities in the photosphere of early-type stars. This
supports a physical mechanism where microturbulence is caused by waves
that are triggered by subsurface convection zones. They further
suggest that clumping in the inner parts of the winds of OB stars
could be related to subsurface convection, and that the convective layers
may also be responsible for stochastic excitation of non-radial
pulsations. Furthermore, magnetic fields produced in the iron convection
zone could appear at the surface of such massive stars. Therefore
subsurface convection could be responsible for the occurrence of
observable phenomena such as line profile variability and discrete
absorption components. These phenomena have been observed for decades,
but still evade a clear theoretical explanation. Here we present
preliminary results from 3D MHD simulations of such subsurface
convection.

\end{abstract}
\firstsection 
\section{Introduction}
Hot luminous stars show a variety of phenomena in their photospheres and in their winds
which still lack a clear physical interpretations at this time. Among these phenomena are
photospheric turbulence, line profile variability (LPVs), discrete absorption components (DACs), wind
clumping, variable or constant non-thermal X-ray and radio emission, chemical composition
anomalies, and intrinsic slow rotation. \citet{2009A&A...499..279C} argued that a convection zone close to the surface of hot, massive stars, could be responsible for various of these phenomena.  This convective zone is caused by a peak in the opacity due to iron recombination and for this reason is referred as the ``iron convection zone'' (FeCZ). A physical connection may exist between
microturbulence in hot star atmospheres and a subsurface FeCZ. 
The strength of the FeCZ is predicted to increase 
with increasing metallicity $Z$,  
decreasing effective temperature $T$ and increasing 
luminosity $L$, and all three predicted trends  
are reflected in observational data of microturbulence obtained in the context of the VLT-FLAMES survey of massive stars \citep{2005Msngr.122...36E}. Moreover recent measurements of microturbulence \citep{2010MNRAS.404.1306F} are in agreement with the results of \citet{2009A&A...499..279C}. 
This suggests that microturbulence corresponds to a physical motion of
the gas in hot star atmospheres. This motion may then be  connected to
wind clumping, since the empirical 
microturbulent velocities are comparable to the local sound speed at the stellar surface. 

The FeCZ in hot stars might also produce
localized surface magnetic fields \citep{2009A&A...499..279C}. Such magnetic fields may become buoyant and reach the surface, creating magnetic spots. 
This could explain the occurrence of DACs (discrete absorption
components in UV absorption lines), also in very hot main sequence stars for which pulsational
instabilities are not predicted. 
Moreover there may be  regions of the upper HR diagram for which the presence of
the FeCZ influences, or even excites, non-radial
stellar pulsations. Interestingly stochastic excitation of non-radial pulsations has been recently found in massive stars \citep{2009Sci...324.1540B,2010A&A...519A..38D}. 

The FeCZ could also turn out to directly affect the evolution of hot massive stars.
If it induces wind clumping, it may alter the stellar wind
mass-loss rate. 
Such a change would also influence the angular
momentum loss. In addition, magnetic fields produced by the iron 
convection zone could lead to an enhanced rate of angular momentum loss.
These effects become weaker for lower metallicity,
where the FeCZ is less prominent or absent. 

Finally, the consequences of the FeCZ might be strongest
in Wolf-Rayet stars. These stars are so hot that the iron opacity peak, 
and therefore the FeCZ, can be directly at the stellar surface,
or --- to be more precise --- at the sonic point of the wind flow \citep{hl96}. 
This may relate to the very strong clumping found observationally
in Wolf-Rayet winds \citep{1999ApJ...514..909L,2006ApJ...639L..75M}, and may be required for an understanding of the
very high mass-loss rates of Wolf-Rayet stars \citep{1995ApJ...448..858E,ki92,hl96}.

\section{Simulations of subsurface convection}\label{mhd}
Convection in the FeCZ is relatively inefficient: the transport of energy is
dominated by radiation, which usually accounts for more than 95\% of
the total flux. This region of the star is very close to the
photosphere, above which strong winds are accelerated. The continuous
loss of mass from the stellar surface moves the convection region to
deeper layers, revealing to surface material that has been processed
in the FeCZ. In rotating stars, the associated angular momentum loss
might also drive strong differential rotation in the region of
interest.
It is clear that, under these circumstances, the mixing length theory
can only give a qualitative picture of the convective properties in
these layers. In order to study the effects induced by the presence of
subsurface convection at the stellar surface in a more quantitative
way, we perform local 3D MHD calculations of
convection. In these simulations we can vary the relative importance
of the background radiative flux and include the effects of rotation
and shear, in order to model conditions as similar as
computationally permitted to subsurface convection in OB stars.

\begin{figure}[b]
\begin{center}
 \includegraphics[width=5.5in]{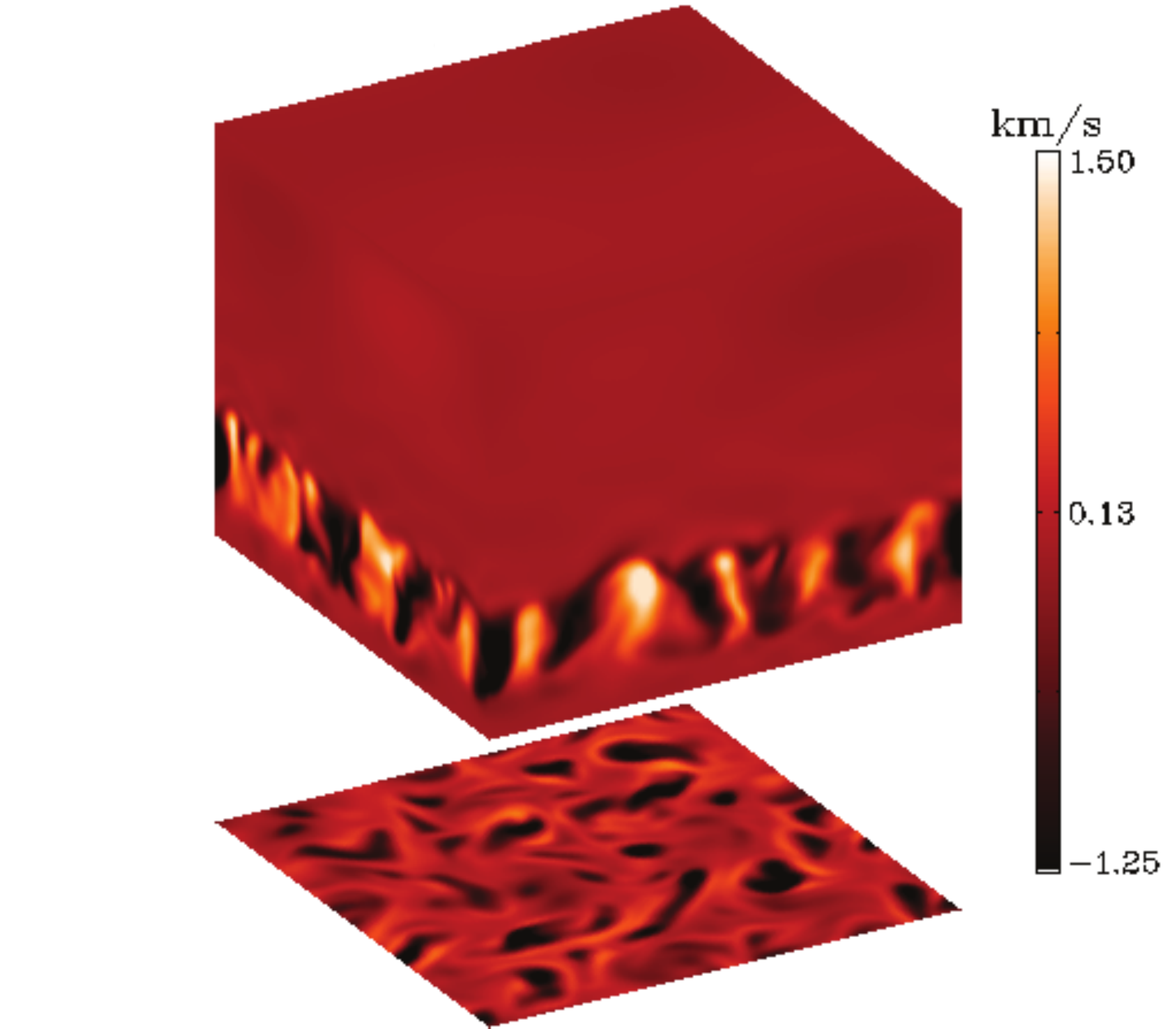} 
 \caption{Simulation of subsurface convection. Starting from the top, the box is divided into
three layers: a radiative layer with an upper cooling boundary, a convectively unstable layer
and another stable layer at the bottom. The snapshot shows values of vertical velocity. 
The plane below the box shows
the vertical velocity field at the lower boundary of the convective layer.}
   \label{fig1}
\end{center}
\end{figure}

\subsection{Computational model}
The setup is similar to that used by \citet{2008A&A...491..353K}.  A rectangular portion of a star is modeled
by a box situated at colatitude $\theta$. The dimensions of the computational domain 
are $(L_x,\, L_y,\, L_z) = (5,\,5,\,5)d$ where $d$ is the depth of the convectively unstable layer.
Our $(x,y,z)$ correspond to $(\theta,\phi,r)$ in spherical polar
coordinates.
The box is divided into
three layers, an upper cooling layer, a convectively unstable layer,
and a stable overshoot layer (see below). The following set of
equations for compressible magnetohydrodynamics is being solved:
\begin{equation}
\frac{\partial \bm A}{\partial t} + Sx \frac{\partial \bm A}{\partial y} 
= {\bm U} \times {\bm B} -\eta\mu_0{\bm J}- S A_y \hat{\bm x}, \label{equ:AA}
 \end{equation}
\begin{equation}
\frac{\mathrm{D} \ln \rho}{\mathrm{D}t} = -\DIV{\bm U},
 \end{equation}
\begin{equation}
 \frac{\mathrm{D} \bm U}{\mathrm{D}t} = -\frac{1}{\rho}{\bm \nabla}p + {\bm g} - 2\bm{\Omega} \times \bm{U} + \frac{1}{\rho} \bm{J} \times {\bm B} + \frac{1}{\rho} \bm{\nabla} \cdot 2 \nu \rho \mbox{\boldmath ${\sf S}$}- S U_x \hat{\bm y}, \label{equ:UU}
 \end{equation}
\begin{equation}
 \frac{\mathrm{D} e}{\mathrm{D}t} = - \frac{p}{\rho}\DIV {\bm U} + \frac{1}{\rho} \bm{\nabla} \cdot K \bm{\nabla}T + 2 \nu \mbox{\boldmath ${\sf S}$}^2 + \frac{\eta}{\rho} \mu_0\bm{J}^2 - \frac{e\!-\!e_0}{\tau(z)}, \label{equ:ene}
 \end{equation}
where $\mathrm{D}/\mathrm{D}t = \pd/\pd t
+ (\bm{U}+\overline{\bm U}_{0}) \cdot \bm{\nabla}$, 
and $\overline{\bm U}_{0} = (0, S\,x,0)$ is an imposed large-scale shear 
flow in the y-direction.
The magnetic field is written in terms of the magnetic vector potential,
$\bm{A}$, with $\bm{B} = \bm{\nabla} \times \bm{A}$,
$\bm{J} =\mu_0^{-1}\bm{\nabla} \times \bm{B}$ is the current density,
$\mu_0$ is the vacuum permeability, $\eta$ and $\nu$ are the magnetic 
diffusivity and kinematic viscosity, respectively, $K$ is the heat 
conductivity, $\rho$ is the density, $\bm{U}$ is the
velocity, $\bm{g} = -g\hat{\bm{z}}$ is the gravitational acceleration,
and $\bm{\Omega}=\Omega_0(-\sin \theta,0,\cos \theta)$ is the rotation vector.
The fluid obeys an ideal gas law $p=(\gamma-1)\rho e$, where $p$
and $e$ are pressure and internal energy, respectively, and
$\gamma = c_{\rm P}/c_{\rm V} = 5/3$ is the ratio of specific heats at
constant pressure and volume, respectively. The specific internal energy per unit mass is related to the
temperature via $e=c_{\rm V} T$.
The rate of strain tensor $\mbox{\boldmath ${\sf S}$}$ is given by
\begin{equation}
{\sf S}_{ij} = \onehalf (U_{i,j}+U_{j,i}) - \onethird \delta_{ij} \DIV \bm{U}.
\end{equation}

The last term of Eq.~(\ref{equ:ene}) describes cooling at the top of
the domain. 
Here $\tau(z)$ is a cooling time which has a profile
smoothly connecting the upper cooling layer and the convectively
unstable layer below, where $\tau\to\infty$.

The positions of the bottom of the box, bottom and top of the
convectively unstable layer, and the top of the box, respectively,
are given by $(z_1, z_2, z_3, z_4) = (-0.5, 0, 1, 4.5)d$, where $d$ is 
the depth of the convectively unstable layer. Initially
the stratification is piecewise polytropic with polytropic indices
$(m_1, m_2, m_3) = (3, 0.9, 3)$, 
which leads to a convectively unstable layer
between two stable layers. 
The cooling 
term leads to a stably stratified isothermal layer at the top.
All simulations with rotation use $\theta=0^\circ$ corresponding to
the north pole.

Stress-free boundary conditions are used in the vertical ($z$) 
direction for the velocity,
\begin{equation}
U_{x,z} = U_{y,z} = U_z = 0,
\end{equation}
where commas denote partial derivatives,
and  perfect conductor conditions are used 
for the magnetic field, i.e.\
\begin{eqnarray}
B_{x,z} = B_{y,z} = B_z &=& 0.
\end{eqnarray}
In the $x$ and $y$ directions
periodic boundary conditions are used. The simulations were performed with
the {\sc Pencil Code}%
\footnote{\texttt{http://pencil-code.googlecode.com/}},
which uses sixth order explicit finite differences in space and third
order accurate time stepping method. Resolutions of up to $128^2\times256$
mesh points were used.

\begin{figure}
 \begin{minipage}{6.85cm}
  \resizebox{\hsize}{!}{\includegraphics{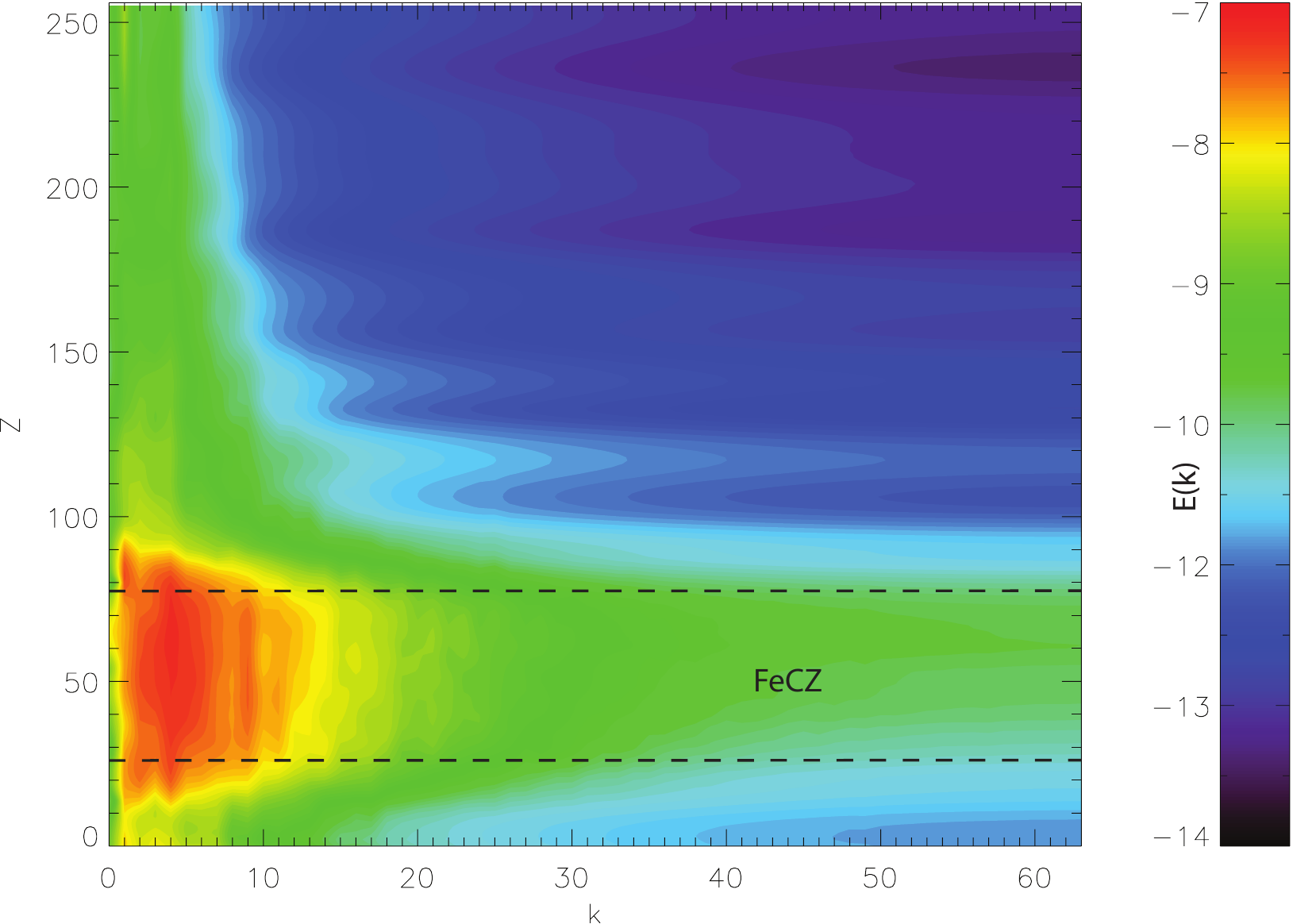}}
 \end{minipage}
 \hfill
 \begin{minipage}{6.6cm} 
   \resizebox{\hsize}{!}{\includegraphics{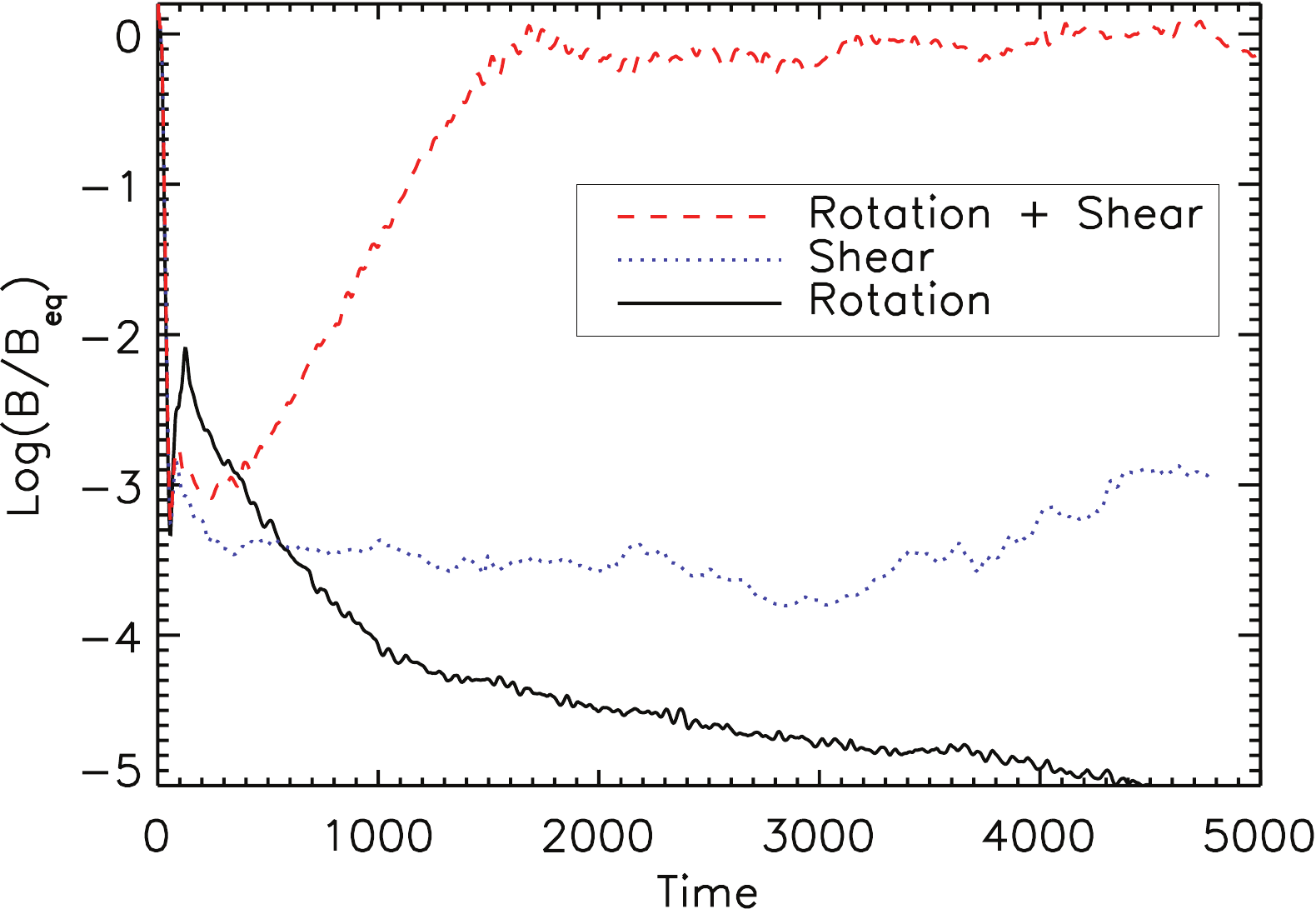}}
 \end{minipage}
\caption{{\bf Left: } evolution of the magnetic field  as function of  time (in code units) in three different runs with different physics. The value on the y-axis is the logarithm of the ratio between the magnetic field and the equipartition value (the value corresponding to equipartition between kinetic and magnetic energy). Dynamo action reaching equipartition is found in simulations including both rotation and shear. {\bf Right: } energy spectrum of the velocity field in our run of subsurface convection. On the $x$-axis is the spatial wavenumber $k$, on the $y$-axis the $z$ coordinate of the  computational domain in gridpoints. The color bar on the right indicates the value of the energy per wavenumber $E(k)$.  The location of the convection zone in the $z$-coordinate is shown by the dashed lines, the rest of the domain is radiative. Just above and below the convection zone kinetic energy is associated with overshooting, while at further distances the transport of energy is facilitated by  gravity waves.}
\label{dynamo}
 \end{figure}

\subsection{Preliminary results}
As a preliminary study we performed low resolution simulations (128x128x256), where the density contrast between the bottom of the convective layer and the top of the domain is only $\sim$20. This is about ten times smaller than in the case of the FeCZ. Moreover the ratio of the convective to radiative flux is about 0.3,  higher than in the FeCZ case. This is because smaller values of convective flux 
result
in steady convection at the low resolution of these preliminary runs.  
Therefore, at this stage, the velocities of convective motions cannot
directly be compared to the velocities obtained by mixing length theory.
However already in these preliminary runs we could follow the excitation
and propagation of gravity waves above the convective region.
In the right panel of Fig.~\ref{dynamo} we show the kinetic energy
spectrum (as function of the spatial wavenumber $k$) in the horizontal
plane,
as function of depth.
The maximum of energy is found in the convective region for those
wavenumbers $k$ corresponding to the number of resolved convective cells
(about 5 along one of the horizontal directions).
Energy is also transported up to the top layer by gravity waves, where
the maximum of the energy is deposited in those wavelengths that are
resonant with the scale of convective motions, as predicted, for example,
by \citet{gk90}.

\citet{2008A&A...491..353K}  found excitation of a large scale dynamo in
simulations of turbulent convection including rotation and shear.
Our computational setup is very similar, so it's not surprising that
we can confirm this result.
Dynamo action reaching equipartition is found in our simulations that
include shear and rotation (see left panel of Fig.~\ref{dynamo}), with
magnetic fields on scales  larger than the scale of convection.

\subsection{Discussion}
The connection found by \citet{2009A&A...499..279C} between the presence of sub-photospheric convective motions and microturbulence in early-type stars is intriguing. The recent identification of solar-like oscillations in hot massive stars \citep{2009Sci...324.1540B,2010A&A...519A..38D} and further measurements of microturbulence \citep{2010MNRAS.404.1306F} also point  toward a picture in which the FeCZ influences surface properties of OB stars. 

We performed 3D MHD simulations of convection to investigate the excitation and propagation of gravity waves above a subsurface convection zone. Analytical predictions of \citet{gk90} on the spatial scale at which the maximum of energy is injected in gravity waves seem to be confirmed in our preliminary calculations. Further investigation is required to understand if the subsurface convection expected in OB stars excites gravity waves of the required amplitude to explain the observed microturbulence in massive stars. In particular we need higher resolution to increase the Reynolds number of our simulations and be able to decrease the ratio of convective to radiative flux, which appear to be an important parameter in determining the convective velocities \citep{2005AN....326..681B}. 

Magnetic fields reaching equipartition values are found in simulations of turbulent  convection if rotation and shear are present (K\"apyl\"a et al. 2008). Since massive stars are usually fast rotators, it could be that the interplay between convection, rotation and shear is able to drive a dynamo in OB stars.  Indeed our simulations of subsurface convection including rotation and shear show the excitation of dynamo action, with magnetic fields  reaching equipartition. This means that fields up to a kG could be present in the FeCZ.  
These magnetic fields might experience buoyant rise and reach the surface of OB stars, where they could have important observational consequences.  In particular it has already been suggested that the discrete absorption components observed in UV lines of massive stars could be produced by low amplitude, small scale magnetic fields at the stellar surface \citep{1994Ap&SS.221..115K}.  
Further study is needed to investigate  the amplitude and geometry of magnetic fields reaching the stellar surface, as well as their effects on the photosphere.

\bibliographystyle{aa} 
\bibliography{ref}
\end{document}